\documentclass[preprint,showpacs,a4paper]{revtex4}
\usepackage{amsmath}
\usepackage{epsf}
\usepackage{epsfig}
\usepackage{graphicx}
\usepackage{amssymb}
\usepackage{fixmath}
\begin{document}

\title{Efficient Bayesian Inference for Learning in the Ising Linear
Perceptron and Signal Detection in CDMA }
\author{Juan P. Neirotti and David Saad}
\affiliation{The Neural Computing Research Group, Aston University,
Birmingham B4 7ET, UK.}

\begin{abstract}
Efficient new Bayesian inference technique is employed for
studying critical properties of the Ising linear perceptron and
for signal detection in Code Division Multiple Access (CDMA). The
approach is based on a recently introduced message passing
technique for densely connected systems. Here we study both
critical and non-critical regimes. Results obtained in the
non-critical regime give rise to a highly efficient signal
detection algorithm in the context of CDMA; while in the critical
regime one observes a first order transition line that ends in a
continuous phase transition point. Finite size effects are also
studied.
\end{abstract}
\maketitle
\section{Introduction}

Efficient inference in large complex systems is a major challenge
with significant implications in science,
engineering and computing.
Exact inference is computationally hard in complex systems and a
range of approximation methods have been devised over the years,
many of which have been originated in the physics literature. A recent
review~\cite{MFAbook} highlights the links between the various
approximation methods and their applications.


In the current paper, we extend a method that was introduced only
recently~\cite{KabashimaCDMA} for inference in dense graphs using message
passing techniques. The method has been employed previously only in the
non-critical regime~\cite{neirottisaad}, and is used here for studying both
critical and non-critical regimes. We apply the method to two different but
related problems: signal detection in CDMA and learning in the Ising linear
perceptron (ILP).

Multiple access communication refers to the transmission of multiple messages
to a single receiver. The scenario we study here is that of $K$ users
transmitting independent messages over an additive white Gaussian noise channel
of zero mean and variance $\sigma_{0}^{2}$. In the
scenario of a Code
Division Multiple Access (CDMA) system~\cite{CDMAbook}, the signal from each
user is modulated by a randomly chosen spreading code of length $N$; these
signals are added up and sent through a noisy channel to the
receiving station, which extracts the original message from the received
signal using knowledge of the user's spreading codes.

We consider the large-system limit, in which the number of users
$K$ tends to infinity while the system load $\beta\equiv K/N\sim\mathcal{O}(1)$.
We focus on a CDMA system using
binary phase shift keying symbols and will assume the power
is completely controlled to unit energy. The received aggregated,
modulated and corrupted signal is then of the form:
\begin{equation}
y_{\mu}=\frac{1}{\sqrt{N}}\sum_{k=1}^{K}s_{\mu
k}b_{k}+\sigma_{0}n_{\mu}\label{eq:CDMA} \ ,
\end{equation}
where $b_{k}$ is the bit transmitted by user $k$, $s_{\mu k}$ the
spreading chip value, $n_{\mu}$  the Gaussian noise variable drawn
from $\mathcal{N}\left(0,1\right)$, and $y_{\mu}$ the received
message. This process is reminiscent of the learning task
performed by a perceptron with binary weights and linear output.

The perceptron 
is a network which
sums a single layer of inputs $s_{\mu j}$, each  weighed by a
corresponding synaptic weight $b_{j}$; the cumulative contribution
is an argument of some transfer function $g(\cdot)$ that gives
rise to the output $y_{\mu}$
\begin{equation}
y_{\mu=}g\left(\frac{1}{\sqrt{K}}\sum_{j=1}^{K}s_{\mu
j}b_{j}\right) \ .\label{eq:perceptron}
\end{equation}
A normalisation factor $K^{-1/2}$ is included to make the argument
of the transfer function of ${\cal O}(1)$. If the entries of 
\textbf{b} are $\pm1$ the perceptron is termed Ising perceptron.
When the transfer function is the identity, the perceptron is
referred to as linear~\cite{Seung}.

The similarity between the linear perceptron of
Eq.~(\ref{eq:perceptron}) and the CDMA detection problem of
Eq.~(\ref{eq:CDMA}) allows for a direct relation between the two
problems to be established. The main difference between the
problems is the regime of interest. While CDMA detection
applications are of interest mainly for non-critical low load
values, ILP studies focused on the critical
regime. We consider both regimes in this paper, but to unify the
treatment we will use the notation and scaling conventions of the
CDMA system.


\section{Message passing}

Graphical models (Bayes belief networks) provide a powerful framework
for modelling statistical dependencies between variables
\cite{Pearl,Jensen,macKay}.  They play an essential role in devising a
principled probabilistic framework for inference in a broad range of
applications.

Message passing techniques are guaranteed to converge to the globally correct
estimate  in graphical models that can be represented by a sparse graph with a
few (typically long) loops. There are no such guarantees for systems with loops
even in the case of large loops and a local tree-like structure (although
see~\cite{weiss}). A clear link has been established between certain message
passing algorithms and methods of statistical
mechanics~\cite{MFAbook,TAPEPL,YFW}.

In a recent development, we presented a new approach~\cite{neirottisaad}, which
was inspired by both, the extension of Belief Propagation (BP)
to tackle densely connected graphs~\cite{KabashimaCDMA} and that
of the replica-symmetric-equivalent BP to Survey Propagation
(SP)~\cite{MPZ}.

The systems we consider here are characterised by multiplicity of
pure states and a possible fragmentation of the space of
solutions. To address the inference problem in such cases we
consider an ensemble of replicated systems where averages are
taken over the ensemble of potential solutions. This amounts to
the presentation of a new graph,
where the observables $y_{\mu}$ are linked to variables in all
replicated systems, namely
$\mathbf{B}\!=\!\left(\mathbf{b}^{1},\mathbf{b}^{2},\dots,\mathbf{b}^{n}\right)$;
where 
$\mathbf{b}^{{\rm
a}}\!=\!\left(b_{1}^{{\rm a}},b_{2}^{{\rm a}},\dots,b_{K}^{{\rm
a}}\right)^{\sf T}$.
To estimate the parameters $\mathbf{B}$ given the data
$\mathbf{y}\!=\!\left(y_{1},y_{2},\ldots,y_{N}\right)^{\sf T}$, in a Bayesian
framework, we have to maximise the posterior
%
$
P\left(\mathbf{B}|\mathbf{y}\right)\!\propto\!
\prod_{\mu=1}^{N}P\left(y_{\mu}|\mathbf{B}\right)P\left(\mathbf{B}\right)
,
$
where we have considered independent data, and thus
$P\left(\mathbf{y}|\mathbf{B}\right)\!=\!\prod_{\mu=1}^{N}
P\left(y_{\mu}|\mathbf{B}\right)$.

The likelihood so defined is of a general form; the explicit
expression depends on the particular problem studied. Here, we are
interested in cases where $\mathbf{b}\! \in \!\left\{ \pm1\right\} ^{K}$
is an unbiased vector and $P\left(\mathbf{B}\right)\!=\!2^{-Kn}$.  The
estimate we would like to obtain is the maximiser of the posterior
marginal (MPM)
$
\widehat{\mathbf{b}}_{k}\!=\!\mathop{\mathrm{argmax}}_{\mathbf{b}_{k}\in\left\{
\pm\right\} ^{n}}\mbox{{\raisebox{-0.8mm}{$
\mathop{\Large\textsf{Tr}}_{\left\{ \mathbf{b}_{l\neq k}\right\}
}$}}}P\left(\mathbf{B}|\mathbf{y}\right) \ ,
$
which is expected to be a vector with equal entries for all
replica
$\widehat{b}_{k}^{1}=\widehat{b}_{k}^{2}=\dots=\widehat{b}_{k}^{n}$.
The number of operations required to obtain the full MPM estimator
is of $\mathcal{O}\left(2^{K}\right)$ which is infeasible for
large $K$ values.

For calculating the posterior in the case of
both CDMA and the ILP, we use the explicit
dependency of $y_{\mu}$ on $\mathbf{b}_{k}$ from Eqs.
(\ref{eq:CDMA}) and (\ref{eq:perceptron})
\(
y_{\mu}\!=\!\sum_{l=1}^{K}\varepsilon_{\mu l}b_{l}+\sigma n_{\mu} \ ,
\)
where $\sigma$ is a free parameter of the model, to be optimised
later in the process; it reflects our ignorance of the true noise
parameter $\sigma_{0}$. The variable $n_{\mu}$ is drawn from ${\cal N}(0,1)$
and the
$\varepsilon_{\mu l}$ are small enough to ensure that
$\sum_{l=1}^{K}\varepsilon_{\mu l}b_{l}\!\sim\!\mathcal{O}(1)$. For
facilitating the derivation that follows we also define the
variable
\(
\mathbf{\Delta}_{\mu} \!\equiv \!\sum_{l=1}^{K}\varepsilon_{\mu
 l}\mathbf{b}_{l} \!= \!\sum_{l\neq k}\varepsilon_{\mu
 l}\mathbf{b}_{l}+\varepsilon_{\mu k}\mathbf{b}_{k} \!=\!
 \mathbf{\Delta}_{\mu k}+\varepsilon_{\mu k}\mathbf{b}_{k} \ ,
\)
representing the uncorrupted signal.
Subsequently, the likelihood can be expanded to take the form

\begin{eqnarray}
P\left(y_{\mu}|\mathbf{B}\right) & \propto &\prod_{{\rm a}=1}^{n}
\exp\left[-\frac{1}{2\sigma^{2}}
\left(y_{\mu}-\Delta_{\mu}^{{\rm a}}\right)^{2}\right]
\label{eq:likelyhood} \\ & \simeq &
\int\prod_{{\rm
a}=1}^{n}\left(\mathrm{d}\Delta_{\mu k}^{{\rm
a}}\exp\left[-\frac{1}{2\sigma^{2}}\left(y_{\mu}-\Delta_{\mu
k}^{{\rm a}}\right)\right]\left[1+\varepsilon_{\mu
k}\frac{\left(y_{\mu}-\Delta_{\mu k}^{{\rm
a}}\right)}{\sigma^{2}}b_{k}^{{\rm
a}}\right]\right)P\left(\mathbf{\Delta}_{\mu k}\right) \ .
\nonumber
\end{eqnarray}

Using Bayes rule one obtains the BP equations
\begin{eqnarray}
P^{t+1}\left(y_{\mu}|\mathbf{b}_{k},\left\{ y_{\nu\neq\mu}\right\}
\right) & = & \mbox{{\raisebox{-1.5mm}{${\displaystyle
\mathop{\LARGE\textsf{Tr}}_{\left\{ \mathbf{b}_{l\neq k}\right\}
}}$}}}P\left(y_{\mu}|\mathbf{B}\right)\prod_{l\neq
k}P^{t}\left(\mathbf{b}_{l}|\left\{ y_{\nu\neq\mu}\right\}
\right)\label{eq:bp1}\\ P^{t}\left(\mathbf{b}_{l}|\left\{
y_{\nu\neq\mu}\right\} \right) & \propto &
\prod_{\nu\neq\mu}P^{t}\left(y_{\nu}|\mathbf{b}_{l},\left\{
y_{\sigma\neq\nu}\right\} \right)\ .\label{eq:bp2}
\end{eqnarray}

An explicit expression for the inter-dependency between solutions
is required for obtaining a closed set of update equations. We
assume a dependence of the form
$
P^{t}\left(\mathbf{b}_{k}|\left\{ y_{\nu\neq\mu}\right\}
\right)\!\propto\!\exp\left\{ \mathbf{h}_{\mu
k}^{t\mathsf{T}}\,\mathbf{b}_{k}+\frac{1}{2}\mathbf{b}_{k}^{\textsf{T}}\mathbf{Q}_{\mu
k}^{t}\,\mathbf{b}_{k}\right\} \ ,
$
where $\mathbf{h}_{\mu k}^{t}$ is a vector representing an external
field and $\mathbf{Q}_{\mu k}^{t}$ the matrix of cross-replica interaction.
We expect the free energy obtained from the well behaved distribution
$P^{t}$ to be self-averaging, thus we assume the following symmetry
between replica
$
\left(\mathbf{Q}_{\mu k}^{t}\right)^{\mathrm{ab}}  =
\left(1-\delta^{\mathrm{ab}}\right)\,g_{\mu
k}^{t}/n$ and $\left(\mathbf{h}_{\mu k}^{t}\right)^{\rm a}  =  h_{\mu
k}^{t},
$
where
both $h_{\mu k}^t$ and $g_{\mu k}^t$ are of ${\cal O}(1)$. An
expression for $P^{t}$ immediately follows
\begin{equation}
P^{t}\left(\mathbf{b}_{k}|\left\{
y_{\nu\neq\mu}\right\} \right)=\frac{{\displaystyle {\displaystyle
{\displaystyle \int_{-\infty}^{\infty}\mathrm{d}x\,\exp{\textstyle
{\displaystyle \left\{ -n\frac{\left(x-h_{\mu
k}^{t}\right)^{2}}{2g_{\mu
k}^{t}}+x\sum_{\textrm{a}=1}^{n}b_{k}^{\textrm{a}}\right\}
}}}}}}{{\displaystyle {\displaystyle
\int_{-\infty}^{\infty}\mathrm{d}x\,\exp\left\{ -n\Phi\left(x;h_{\mu
k}^{t},g_{\mu k}^{t}\right)\right\} }}},\label{pp}\end{equation} where
$ \Phi\left(x;h_{\mu k}^{t},g_{\mu
k}^{t}\right)\!=\!\left(x-h_{\mu k}^{t}\right)^{2}/2g_{\mu
k}^{t}-\ln\left(2\cosh(x)\right) .
$
We exploit the assumption that the number of replica $n$ is large
and employ Laplace's method to find dominant contributions to the
integral. The function $\Phi\left(x;h_{\mu k}^{t},g_{\mu
k}^{t}\right)$ exhibits two minima if $h_{\mu k}^{t}\to0$ and
$g_{\mu k}^{t}>1$; these will provide the only contributions in
that limit. Other regimes will provide trivial solutions. If
the field $h_{\mu k}^{t}$ goes to zero as
\( m_{\mu k}^{t}h_{\mu k}^{t}\!\sim\!\ln\left(4n\left(n_{\mu
k}^{t}\right)^{-2}\right)/2n \ ,
\)
where $m_{\mu k}^{t}$ is the spontaneous magnetisation and $n_{\mu
k}^{t}$ a constant, the first two moments of $b_{k}^{\textrm{a}}$ are, up to
${\cal O}\left(n^{-1}\right)$,
\begin{eqnarray}
\left\langle b_{k}^{\textrm{a}}\right\rangle &=&
 \mbox{{\raisebox{-0.75mm}{${\displaystyle
 \mathop{\LARGE\textsf{Tr}}_{\left\{ \mathbf{b}_{k}\right\}
 }}$}}}P^{t}\left(\mathbf{b}_{k}|\left\{ y_{\nu\neq\mu}\right\}
 \right)b_{k}^{\textrm{a}} \;\simeq\;
 \left[1-\frac{\left(n_{\mu k}^{t}\right)^{2}}{2n}\right]m_{\mu
 k}^{t}\nonumber\\ 
 \left\langle
 b_{k}^{\textrm{a}}b_{l}^{\textrm{b}}\right\rangle - \left\langle
 b_{k}^{\textrm{a}}\right\rangle \left\langle
 b_{l}^{\textrm{b}}\right\rangle &\simeq& \delta_{kl}\left\{ \delta^{{\rm
 ab}}\left[1-\left(m_{\mu
 k}^{t}\right)^{2}\right]+\left(1-\delta^{{\rm
 ab}}\right)\frac{\left(n_{\mu k}^{t}m_{\mu
 k}^{t}\right)^{2}}{n}\right\} \ .\nonumber
\end{eqnarray}

However, for calculating the posterior we need the distribution on
the variable $\Delta_{\mu k}^{\rm a}$, which is a sum of a large number of
unbiased and uncorrelated random variables $\varepsilon_{\mu k}$
and $b_{k}^{\mathrm{a}}$. Therefore, by virtue of the central
limit theorem, the variable $\Delta_{\mu
k}^{\mathrm{a}}=\sum_{l\neq k}\varepsilon_{\mu
l}b_{l}^{\mathrm{a}}$ obeys a normal distribution, whose mean
value and covariance matrix are given by
\begin{eqnarray}
\left(\mathbf{u}_{\mu k}^{t}\right)^{\mathrm{a}}\equiv
\left\langle \Delta_{\mu k}^{\mathrm{a}}\right\rangle  & = &
\mathbf{\mbox{{\raisebox{-1.5mm}{${\displaystyle
\mathop{\LARGE\textsf{Tr}}_{\left\{ \mathbf{b}_{l\neq k}\right\}
}}$}}}} \prod_{l\neq k}P^{t}\left(\mathbf{b}_{l}|\left\{
y_{\nu\neq\mu}\right\} \right)\sum_{l\neq k}\varepsilon_{\mu
l}b_{l}^{\mathrm{a}}
 =  \sum_{l\neq k}\varepsilon_{\mu l}m_{\mu l}^{t}\label{eq:meanu}\\
\left(\mathbf{{\Upsilon}}_{\mu k}^{t}\right)^{\mathrm{ab}}\equiv\left\langle
\Delta_{\mu k}^{\textrm{a}}\Delta_{\mu k}^{\textrm{b}}\right\rangle -
\left\langle \Delta_{k}^{\textrm{a}}\right\rangle \left\langle
\Delta_{k}^{\textrm{b}}\right\rangle  & = & \mathbf{\mbox{{\raisebox{-1.5mm}
{${\displaystyle \mathop{\LARGE\textsf{Tr}}_{\left\{ \mathbf{b}_{l\neq k}
\right\} }}$}}}}\prod_{l\neq k}P^{t}\left(\mathbf{b}_{l}|
\left\{ y_{\nu\neq\mu}\right\} \right)\sum_{\substack{l\neq k\\
j\neq k}
}\varepsilon_{\mu l}\varepsilon_{\mu j}b_{l}^{\mathrm{a}}b_{j}^{\mathrm{b}}
-\left(\sum_{l\neq k}\varepsilon_{\mu l}m_{\mu l}^{t}\right)^{2}\nonumber \\
 =  \sum_{l\neq k}\varepsilon_{\mu l}^{2}
 \left\{ \left\langle b_{l}^{\mathrm{a}}b_{j}^{\mathrm{b}}\right\rangle
 -\left\langle b_{l}^{\textrm{a}}\right\rangle \left\langle b_{j}^{\textrm{b}}
 \right\rangle \right\}
 & = & \delta^{\mathrm{ab}}\left(X_{\mu k}-Q_{\mu k}^{t}\right)-
 \left(1-\delta^{\mathrm{ab}}\right)\frac{1}{n}R_{\mu k}^{t}\ ,\label{eq:covmac}
 \end{eqnarray}
where $X_{\mu k}\equiv\sum_{l\neq k}\varepsilon_{\mu l}^{2}$,
$Q_{\mu k}^{t}\equiv\sum_{l\neq k}\left(\varepsilon_{\mu l}m_{\mu
l}^{t}\right)^{2}$ and $R_{\mu k}^{t}\equiv\sum_{l\neq
k}\left(\varepsilon_{\mu l}n_{\mu l}^{t}m_{\mu l}^{t} \right)^{2}$
are macroscopic variables of $\mathcal{O}(1)$. In particular,
$R_{\mu k}^{t}$ is a free variable that can be used to optimise
with respect to a given performance measure. This corresponds to
fine tuning of the variational model considered. All three
quantities $X_{\mu k}$, $Q_{\mu k}^{t}$ and $R_{\mu k}^{t}$ are
self averaging so we can drop both indices $\mu$ and $k.$
The probability of $\mathbf{\Delta}_{\mu k}$ can be expressed as
$
P\left(\mathbf{{\Delta}}_{\mu
k}\right)\propto \exp\left\{
-\frac{1}{2}\left(\mathbf{{\Delta}}_{\mu k}-\mathbf{u}_{\mu
k}^{t}\right)^{\mathsf{T}}\left(\mathbf{{\Upsilon}}_{\mu
k}^{t}\right)^{-1}\left(\mathbf{{\Delta}}_{\mu k}-\mathbf{u}_{\mu
k}^{t}\right)\right\} \ .
$

Having the probability distribution of $\mathbf{\Delta}_{\mu k}$
we can express the message from nodes $y_{\mu}$ to nodes
$b_{k}^{{\rm a}}$ at time $t+1$ explicitly, using
Eqs.~(\ref{eq:likelyhood}), (\ref{eq:bp1}) and (\ref{eq:bp2})
\begin{equation}
\widehat{m}_{\mu k}^{t+1} =  \frac{\mathbf{\mbox{{\raisebox{-0.75mm}
{$\mathop{\LARGE\textsf{Tr}}_{\left\{
\mathbf{b}_{k}\right\} }$}}}}\; b_{k}^{\widetilde{{\rm a}}}\,
P^{t+1}\left(y_{\mu}|\mathbf{b}_{k}, \left\{
y_{\nu\neq\mu}\right\} \right)}{\mathbf{\mbox{{\raisebox{-0.75mm}
{$ \mathop{\LARGE\textsf{Tr}}_{\left\{
\mathbf{b}_{k}\right\} }$}}}}\;
P^{t+1}\left(y_{\mu}|\mathbf{b}_{k}, \left\{
y_{\nu\neq\mu}\right\} \right)}
 =  {\displaystyle \frac{\varepsilon_{\mu k}}{\sigma^{2}+\beta-Q^{t}+R^{t}}}
 {\displaystyle \left(y_{\mu}-u_{\mu k}^{t}\right)} \ .
 \end{equation}

It is then straightforward to prove the following equation and its approximation,
the latter due to the fact that $\widehat{m}_{\nu
k}^{t}\sim\mathcal{O}\left(\varepsilon_{\nu k}\right)$
\begin{equation} m_{\mu
k}^{t}=\tanh\left(\sum_{\nu\neq\mu}{\rm
arctanh}\left(\widehat{m}_{\nu k}^{t}\right)\right)
\simeq\tanh\left(\sum_{\nu\neq\mu}\widehat{m}_{\nu k}^{t}\right) \
. \label{eq:mmhatapprox}
\end{equation}

To study the quality of the inferred vectors one considers the
gauged field with respect to the true message $b_{k}h_{\mu k}^{t}$
where $h_{\mu k}^{t}\equiv{\rm artanh}\left(m_{\mu
k}^{t}\right)=\sum_{\nu\neq\mu}{\rm artanh}\left(\hat{m}_{\nu
k}^{t}\right)\simeq\sum_{\nu\neq\mu}\hat{m}_{\nu k}^{t}$. The
distribution of this field is likely to be well approximated by a
Gaussian, as a result of the central limit theorem, whose mean and
variance  are $E^{t}$ and $F^{t}$ respectively
\begin{equation}
E^{t}  =  \frac{1}{K}\sum_{k=1}^{K}\sum_{\mu=1}^{N}b_{k}\hat{m}_{\mu k}^{t}\,,
\qquad
F^{t}   \simeq
\frac{1}{K}\sum_{k=1}^{K}\sum_{\mu=1}^{N}\left(\hat{m}_{\mu
k}^{t}\right)^{2}\, ;\label{eq:fdet}
\end{equation}
%
%
%
both are assumed to be independent of the index $\mu$ due to
self-averaging. For the same reason we expect the macroscopic
variables, representing the overlap between the vectors
$\mathbf{m}_{\mu}$ and $\mathbf{b}$ at any time $t$ and the
squared length of $\mathbf{m}_{\mu}$, defined as
$M_{\mu}^{t}\equiv\sum_{k=1}^{K}b_{k}m_{\mu
k}^{t}/K\simeq\sum_{k=1}^{K}b_{k}m_{k}^{t}/K=M^{t}$ and
$N_{\mu}^{t}\equiv\sum_{k=1}^{K}\left(m_{\mu
k}^{t}\right)^{2}/K\simeq\sum_{k=1}^{K}\left(m_{k}^{t}\right)^{2}/K=N^{t}$,
to be $\mu$ independent. Using the distribution we obtained for
the gauged field $b_{k}h_{\mu k}^{t}$, both variables can be
evaluated by
$
M^{t}  =  \int\mathcal{D}u\,\tanh\left(\sqrt{F^{t}}u+E^{t}\right),$ 
$N^{t}  = 
\int\mathcal{D}u\,\tanh^{2}\left(\sqrt{F^{t}}u+E^{t}\right)
,
$
where $\mathcal{D}u=\exp\left(-u^{2}/2\right)/\sqrt{2\pi}$.
Applying a method equivalent to the EM algorithm~\cite{EM} for the
independent parameter of the model $\sigma^2$ we have that the
optimal selection of the parameter is given by the condition
$E^{t}=F^{t}$, which also implies that $N^{t}=M^{t}$. Notice that
this result is not surprising as it maximises the normalised
overlap between the vectors $\mathbf{m}_{\mu}$ and $\mathbf{b}$.

It is important to notice at this point the different scaling
factors used in the two models we examine.  For CDMA one uses
$\varepsilon_{\mu k}=s_{\mu k}/\sqrt{N}$ while $\varepsilon_{\mu
k}=s_{\mu k}/\sqrt{K}$ is used for the
(ILP). Imposing the condition $E^{t}=F^{t}$ leads to a relation
between the structure of the space of solutions, represented by
$R^{t}$, and the free parameter of the model $\sigma^{2}$. From
Eqs.~(\ref{eq:fdet}) one obtains for the two
models
$
E^{t+1}  =  {\rm e}_1^{-1} \left[\sigma^{2}+R^{t}+{\rm
e}_2\left(1-N^t\right)\right]^{-1}\,,$
$F^{t+1}  =  {\rm e}_1 \left[\sigma^{2}_0+{\rm e}_2\left(1-N^t\right)\right]\left(E^{t+1}\right)^2\;,
$
where ${\rm e}_1=1\;(\beta)$ and ${\rm e}_2=\beta\;(1)$ for the CDMA (ILP)
system. This implies, after simplification, that for both cases
$R^{t}=\sigma_{0}^{2}-\sigma^{2}$.

Despite the simplicity of this result, the process from which we
obtained it provides a mechanism for estimating the true noise
variance. In deriving $E^{t}$ and $F^{t}$ we used the fact that
$K,N\to\infty\,{\rm with}\ K/N=\beta$. So that the true noise
variance $\sigma_{0}^{2}$ that appears in the expression for
$F^{t}$ has been obtained from a signal vector with an infinite
number of entries $y_{\mu}$. Thus 
\(
\lim_{N\to\infty}\frac{1}{N}\sum_{\mu=1}^{N}\left(y_{\mu}\right)^{2}={\rm e}_2+\sigma_{0}^{2}\,.
\)
Using this we can express the message
as
\begin{equation}
\widehat{m}_{\mu k}^{t+1} \simeq \varepsilon_{\mu k}  \left[{\displaystyle \frac{1}{N}\sum_{\mu=1}^{N}
 \left(y_{\mu}\right)^{2}}-{\rm e}_2N^{t}\right]^{-1}\left(y_{\mu}-u_{\mu k}^{t}\right)\, ,
 \label{mhatfin}
\end{equation}
where no prior belief of the noise level $\sigma_{0}$ is required.

The steady state equation for the macroscopic variable 
$E^{t}$ is obtained in the limit $t\to\infty$, leading to the
definition of
$\overline{E}\equiv \lim_{t\to\infty}E^{t}$. In this 
regime the following relation holds
\begin{equation}
\overline{E}\left(\sigma_{0}^{2},\beta\right) =  {\rm e}_1^{-1}
\left\{\sigma^{2}_0+{\rm e}_2\left[1-\int\mathcal{D}u\,\tanh^{2}
\left(\sqrt{\overline{E}\left(\sigma_{0}^{2},\beta\right)}u+\overline{E}
\left(\sigma_{0}^{2},\beta\right)\right)\right]\right\}^{-1} \ .
\label{eq:ebar}
\end{equation}
%
%
%
From these expressions one can calculate directly the error per
bit rate
\begin{equation}
\overline{P}_{b}\left(\sigma_{0}^{2},\beta\right)=\frac12\left(1+{\rm
erf}\left(\sqrt{\frac{\overline{E}\left(\sigma_{0}^{2},\beta\right)}{2}}\right)\right)\,.
\label{eq:epbbar}
\end{equation}
%

\begin{figure}[t]
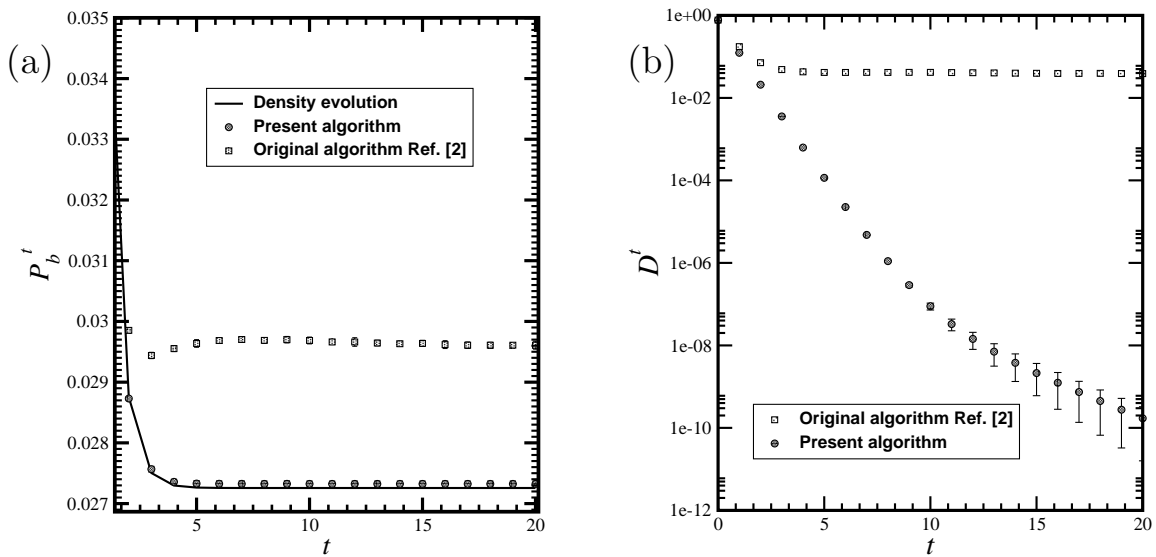

\begin{picture}(440,205)
\put(5,0){\epsfxsize=70mm  \epsfbox{plot1a.eps}}
\put(235,0){\epsfxsize=70mm  \epsfbox{plot2a.eps}}
\put(0,183){\large (a)} \put(235,183){\large (b)}
\end{picture}
\caption{(a) Error probability of the inferred solution as a
function of time.  The system load $\beta\!=\!0.25$, true and
estimated noise levels \( \sigma _{0}^2 \!= \!0.25 \) and \(
\sigma^2 \!=\! 0.01 \), respectively. Squares represent results
obtained by the algorithm of~\cite{KabashimaCDMA},
solid line the dynamics obtained from our equations; circles
represent results obtained from the suggested {\em practical}
algorithm. Variances are smaller than the symbol size.  (b) The
measure of convergence $D$ of the obtained solutions, as a
function of time; symbols are as in (a).} \label{fig3}
\end{figure}

\section{Numerical results}
The inference algorithm requires an iterative update of
Eqs.~(\ref{eq:mmhatapprox}) and (\ref{mhatfin}) until they converge to a
reliable estimate of the signal. We emphasise again that there is
no need for prior information on the noise level. The
computational complexity of the algorithm, as it has been
presented here, is of ${\mathcal{O}}(NK^{2})$ but can be reduced
to be ${\mathcal{O}}(K^{2})$ in a similar way to the approach
taken in~\cite{KabashimaCDMA}.
\begin{figure}[t]
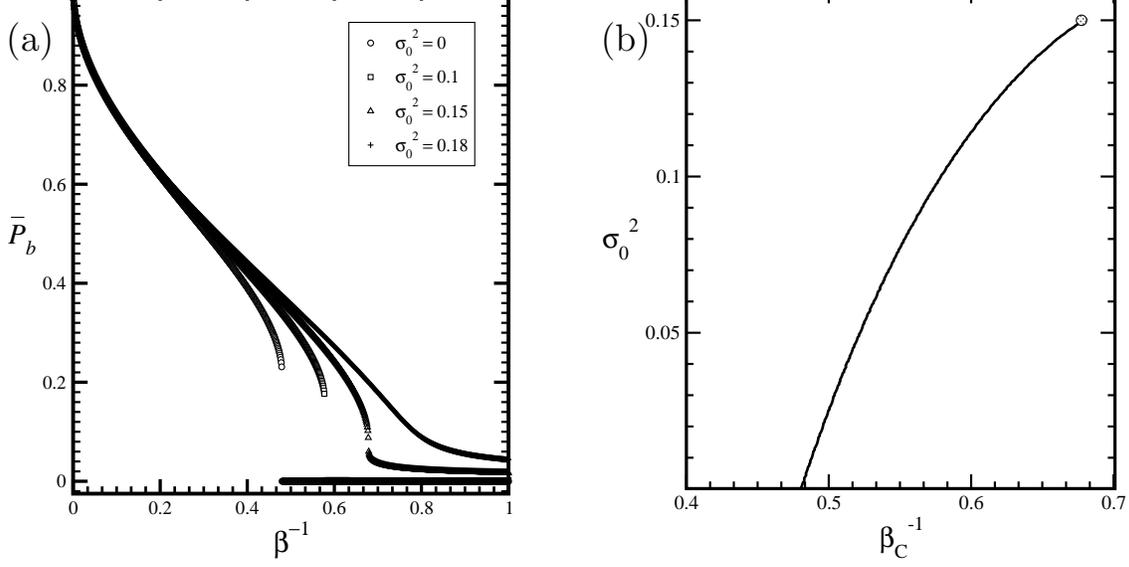

\begin{center}
\begin{picture}(440,190)
\put(0,0){\epsfxsize=67.5mm  \epsfbox{critico1.eps}}
\put(225,0){\epsfxsize=70mm \epsfbox{critico2.eps}}
\put(0,193){\large (a)} \put(225,193){\large (b)}
\end{picture}
\end{center}
\caption{(a) $\overline{P}_b$ at the steady state,
Eq.~(\ref{eq:epbbar}), as a function of $\beta^{-1}$ for
different values of the noise parameter. For values of $\sigma_{0}^{2}$
below 0.15 the curves show
discontinuity at certain $\beta$ values, which becomes continuous
but non-analytic at $\sigma_{0}^{2}=0.15$ around
$\beta^{-1}\simeq0.68$. For noise variance values above
$\sigma_{0}^{2}=0.15$ the curves become
analytical. (b) Position of the non analyticity of the error rate
curve $\beta_{C}^{-1}$ as a function of the noise
parameter $\sigma_{0}^{2}$. This first order phase
transition-like curve ends in a second order phase transition-like
point marked by $\circ$.} \label{fig4}
\end{figure}

To test the performance of our algorithm we carried out a set of
experiments of CDMA signal detection under typical conditions.
Error probability of the inferred signals has been calculated for
a system load of $\beta\!=\!0.25$, where the true noise level is
$\sigma_{0}^{2}\!=\!0.25$ and the estimated noise is
$\sigma^{2}\!=\!0.01$, as shown in Figure~\ref{fig3}(a). The solid
line represents the expected theoretical results (density
evolution), knowing the exact values of $\sigma_{0}^{2}$ and
$\sigma^{2}$, while circles represent simulation results obtained
via the suggested \emph{practical} algorithm, where no such
knowledge is assumed. The results presented are based on $10^{5}$
trials per point and a system size $N\!=\!2000$, and are superior
to those obtained using the original algorithm of
Ref.~\cite{KabashimaCDMA}.

Another performance measure to be consider is \(
D^{t}\!\equiv\! K^{-1}\!\left(\mathbf{m}^{t}-\mathbf{m}^{t-1}\right)\cdot
\left(\mathbf{m}^{t}-\mathbf{m}^{t-1}\right),\)
that provides an indication to the stability of the solutions
obtained. In Fig.~\ref{fig3}(b) we compare results obtained from
our algorithm, that exhibit fast convergence to a reliable
solution, in stark contrast to the original
algorithm~\cite{KabashimaCDMA} which does not converge. 

For the ILP, the $K\!>\!N$ regime is highly
interesting as the system develops a critical behaviour for a
range of ($\sigma_{0}^{2}$) values. We carried
out a set of experiments for this system (the CDMA scaling was
kept for consistency) based on density evolution. In
Fig.~\ref{fig4}(a) we present curves of 
$\overline{P}_b$, defined in Eq.~(\ref{eq:epbbar}), as a function of
the inverse load $\beta^{-1}$ for different values of 
$\sigma_{0}^{2}$. Three different regimes have been
observed: For $\sigma_{0}^{2}<0.15$ the curves exhibit a
discontinuity at a value of $\beta$ that varies with
$\sigma_{0}^{2}$ (first order phase transition-like behaviour). At
$\sigma_{0}^{2}\!=\!0.15$ the curve becomes continuous but its slope
diverges (second order phase transition-like behaviour). The 
$\overline{P}_b$ curves show analytical behaviour for noise values
above 0.15.
In Fig.~\ref{fig4}b we present a phase diagram of the CDMA
system. It shows the dependency of the critical load
$\beta_{C}^{-1}$ as a function of the noise parameter. The first
order line ends in a second order transition point marked by a
circle.

\begin{figure}[t]
\begin{center}
\begin{picture}(440,190)
\put(0,-4){\epsfxsize=69.5mm  \epsfbox{critico3.eps}}
\put(225,-5){\epsfxsize=76mm \epsfbox{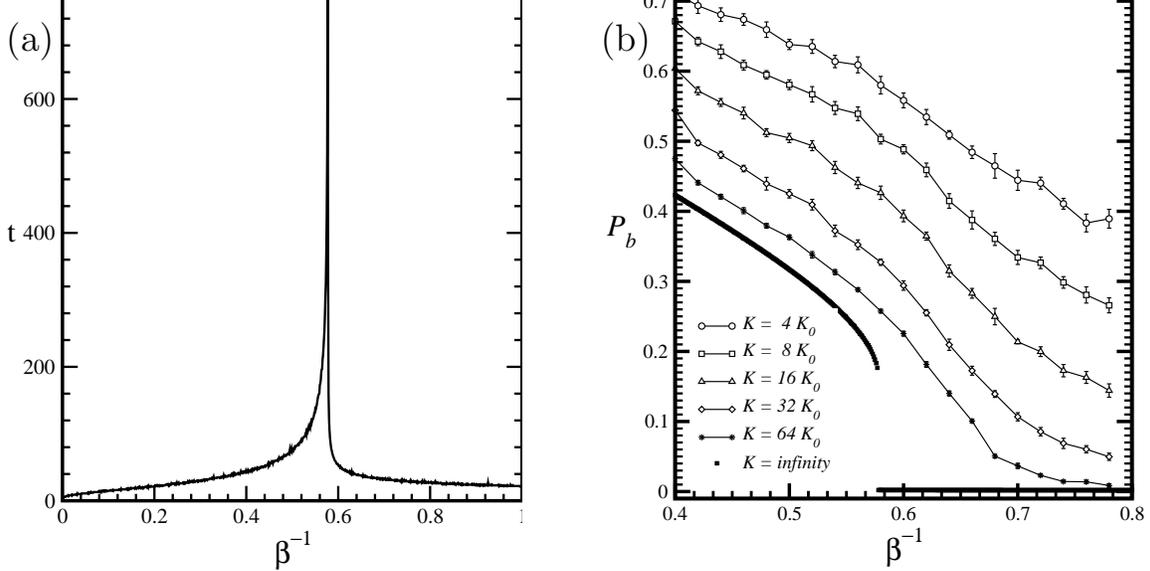}}
\put(0,193){\large (a)} \put(225,193){\large (b)}
\end{picture}
\end{center}
\caption{(a) Number of iterations of Eq.~(\ref{eq:ebar}) required for convergence as a function of $\beta$,
for $\sigma_{0}^{2}=0.10$. 
the error rate curve exhibits a discontinuity. 
(b) Finite size
effects observed in the error rate curve when the 
Eqs.~(\ref{eq:mmhatapprox}) and (\ref{mhatfin}) are iterated
over the number of steps needed to reach the steady state. The
noise level used is $\sigma_{0}^{2}=0.10$ with  $K_{0}=50$. 
The curves are mean values over
1000 experiments.
The curve obtained from the iteration of the steady state
equations is presented as a reference.} \label{fig5}
\end{figure}

Another indication for the critical behaviour
is the number of steps required for the recursive
update of Eq.~(\ref{eq:ebar}) to convergence.
In Fig.~\ref{fig5}(a) we present the number of iterations needed
to reach a steady state as a function of $\beta^{-1}$ when the
noise parameter is set to $\sigma_{0}^{2}=0.10$. The number of
iterations diverge when the critical value of $\beta$ is reached.

Finally, we wish to explore the efficiency of the algorithm as a
function of the system size. In Fig.~\ref{fig5}(b) we present the
result of iterating Eqs.~(\ref{eq:mmhatapprox}) and
(\ref{mhatfin}) for system sizes of $K=$200, 400,
800, 1600 and 3200. The curves represent mean
values over 1000 experiments. There is a strong dependency of the error per
bit rate on the size of the system, which is expected to converge
to the asymptotic limit (infinite system size) represented by the
solid line.

\section{Conclusions}
In summary, we employed a new variational algorithm based on
replicated variable systems to investigate two related problems:
signal detection in CDMA  and learning in the ILP. The new algorithm facilitates the use of message
passing techniques in densely connected systems, even in systems
that show a fragmented solution space and represents an extension
of existing algorithms similar to the extension of BP to SP.

Results on the CDMA signal detection problem are superior than other existing
algorithms~\cite{KabashimaCDMA,kk}, without using any prior for the expected
noise level.

Results have also been obtained for low and intermediate load
levels under various noise conditions, which are of higher
relevance to ILP learning than to CDMA. These
exhibit a first-order like transition for critical load levels and
below a certain noise level ($\sigma_{0}^{2}< 0.15$), that become
second order as the noise level increases (at
$\sigma_{0}^{2}=0.15$). No transition points have been identified
above this noise level. Finally, we also examined finite size
effects in the system, which are clearly present even at a system
size of 3200 nodes.

We are in the process of examining the suitability of the method
for other applications~\cite{neirottisaad_long}. While the
approach seems promising, there is clearly a need for further
research to fully determine the potential of the new algorithm.

\begin{acknowledgements}
Support from EVERGROW IP No.~1935 in FP6 is gratefully
acknowledged.
\end{acknowledgements}


\begin{thebibliography}{99}
\bibitem{kk}Comparison to
the recently presented signal detection algorithm of
Ref.~\cite{Kabashima_new} cannot be carried out due to absence of
numerical data.
\end{thebibliography}
\end{document}